\documentclass[prb,twocolumn,superscriptaddress,showpacs,amsmath,amssymb]{revtex4}

\usepackage{graphicx}
\usepackage{latexsym}
\usepackage{amsmath}
\usepackage{amssymb}
\usepackage{amsfonts}
\usepackage{color}
\usepackage{bm}
\usepackage{verbatim}
\usepackage{color}

\begin{document}

\title{Microscopic derivation of two-component Ginzburg-Landau model
and conditions of its applicability in two-band systems}

\author{Mihail Silaev}
\affiliation{Department of Theoretical Physics, The Royal
Institute of Technology, Stockholm, SE-10691 Sweden}
 \affiliation{Institute for Physics of Microstructures RAS, 603950
Nizhny Novgorod, Russia.}

\author{Egor Babaev}%
\affiliation{Department of Theoretical Physics, The Royal
Institute of Technology, Stockholm, SE-10691 Sweden}
 \affiliation{ Department
of Physics, University of Massachusetts Amherst, MA 01003 USA }
\begin{abstract}
 We report a microscopic derivation of two-component
Ginzburg-Landau (GL) field theory and
the conditions of its validity in two-band superconductors.
We also investigate the conditions when microscopically derived or
phenomenological  GL models fail and one should resort to a  microscopic description.
 We show that besides being directly applicable
at elevated temperatures,  a version of a minimal two-component GL
theory in certain cases also gives  accurate description of
certain aspects of a two-band system even substantially far from
$T_c$. This shows that two-component GL model can be used for
addressing a wide range of questions in multiband systems, in
particular vortex physics and magnetic response. We also argue
that single Ginzburg-Landau parameter cannot in general
characterize magnetic response of multiband systems.
\end{abstract}

 \maketitle
\section{Introduction} Ginzburg-Landau theory of single-component superconductors
has historically proven its strong predicting power and its
extraordinary value as a phenomenological tool. This is despite
the fact that {\it formally} it can be justified only in some
cases in a very narrow band  of temperatures. The temperature  on
the one hand should be high enough to permit an expansion in a
small order parameter. On the other hand  the temperature should
not be too high because the mean-field theory becomes invalid near
$T_c$ due to { critical} fluctuations. Nonetheless the great
success of the GL theory is due to the fact that { it yields} a
qualitatively correct picture in extremely wide range of
temperatures even when its application cannot be justified on
formal grounds.

Shortly after the theoretical proposal of two-band
superconductivity and more recently, two-component GL (TCGL)
expansions were done in application to two-band systems, see e.g.
\cite{Dao}. However in contrast to single-component GL theory, the
conditions under which  TCGL model is valid is  still widely
believed to be  an open question.
 In this work, we resolve this question.
 By taking advantage of the recently calculated normal modes and
length scales in two-band Eilenberger model \cite{Silaev} we
present the first self-consistent microscopic analysis of the
applicability of TCGL theories to describe both linear and
nonlinear responses of two-component superconductors. The results
validate applicability to TCGL for studying of a wide spectrum of
physical questions, including aspects of physics in
low-temperatures regimes.

\textbf{ The problem:} the temperature range of validity of TCGL
model is  bounded from below by a requirement that the field
amplitudes should be small.
 More important bound follows  from the
  observed, under certain conditions, disappearance of one of two  fundamental length scales
 governing the asymptotical behavior of the superfluid density in microscopic
 theories\cite{Silaev}. This implies  that classical two-component field theory obtained
in power-law expansion does fail at low temperatures or at
substantially strong interband couplings \cite{Silaev}. Also, like
for its single-component counterpart, the region of validity of
TCGL expansion is bounded from above by a fluctuation region.

 The key difference between TCGL and single-component GL
theory is the fact that the former  has several coherence lengths.
The existence of multiple length scales which arise from
hybridized normal modes of the linearized TCGL theory, can
dramatically affect the magnetic response of the system
\cite{GL2mass}.
 Under certain conditions it results in situations
where the London penetration length falls between two coherence
lengths \cite{bs1}, which was recently termed ``type-1.5" regime
\cite{moshchalkov} (for a recent brief review see \cite{review}).
However because  the condensates in the bands are not
independently conserved, in the limit  $T \to T_c$ there should be
indeed only  one divergent length scale associated with density
variations. This  in turn implies that in certain cases there also
could be a temperature range close to $T_c$ where long-wavelength
physics is well approximated by a single-component GL theory
(although this regime is not generic since its width is controlled
by different parameters than the fluctuation region and thus  it
can be non-existent because of critical fluctuations). { Therefore
the conditions of the applicability of TCGL are quite different
from that of single-component GL theory and warrant a careful
investigation}, which we present below.

\section{Model and basic equations}

Expansion in powers of gradients and gap
functions of  microscopic equations yields the two-component
Ginzburg-Landau (TCGL) free energy density:
\begin{eqnarray}
 F &=& \Big\{ \sum_{j=1,2}\left(a_j | \Delta_j |^2
 + \frac{b_j}{2}| \Delta_j|^4 + K_j |\bm D\Delta_j |^2\right)   \nonumber\\
 & -&   \gamma \left( \Delta_1\Delta_2^*+\Delta_2\Delta_1^*\right)+\frac{ B ^2}{8\pi } \Big \}\,.
\label{Gibbs}
\end{eqnarray}
 where ${\bm D}={\bm \nabla}  + i {\bm A}$, ${\bm A}$ and ${\bm
B}$ are the vector potential and magnetic field
 and  $\Delta_{1,2}$ are the gap functions in two different
 bands. Despite the fields $\Delta_{1,2}$ are  often
 called ``two order parameters" in the literature, below we
 avoid this terminology since it is not quite accurate.
First there is only  $U(1)$ local symmetry in this model
inspite of the presence of two components,
 since the other global $U(1)$ symmetry is explicitly broken by the terms  $\gamma \left( \Delta_1\Delta_2^*+\Delta_2\Delta_1^*\right)$.
Second, and more important circumstance is that the applicability
of the Ginzburg-Landau or Gross-Pitaevskii classical field theory
does not in general require any broken symmetries. The simplest
example being two-dimensional superfluids at finite temperature:
they can be indeed be described by Gross-Pitaevskii classical
complex field, yet they do not possess spontaneously broken
symmetry. Likewise in superfluid turbulence there is even no
algebraic long-range order yet the system can be described by a
classical complex field. Indeed in some cases, like e.g. in
$U(1)\times U(1)$ superconductors or superfluids one can write
down two-component classical field theory on symmetry grounds. A
$U(1)$ system such as two-band superconductors can also under
certain conditions be described by two-component classical field
theory, although it does not automatically follows from its
symmetry. The main aim of this paper is to analyze under which
conditions two-band superconductors are described by TCGL theory.

 To verify applicability of TCGL theory we present a comparative study
 of linear response and non-linear regime in TCGL and exact microscopic theories.
 We consider the microscopic model of clean superconductor with two
overlapping bands at the Fermi level \cite{Silaev}. Within
quasiclassical approximation the band parameters characterizing
the two different cylindrical sheets of the Fermi surface are the
Fermi velocities $V_{Fj}$ and the partial densities of states
(DOS) $\nu_j$, labelled by the band index $j=1,2$.

It is convenient to normalize the energies to the critical
temperature $T_c$ and length to $r_0= \hbar V_{F1}/T_c$. The
vector potential is normalized by $\phi_0 /(2\pi r_0)$, the
current density normalized by $c\phi_0/(8\pi^2 r_0^3)$ and
therefore the magnetic field is measured in units
 $\phi_0 /(2\pi r^2_0)$ where $\phi_0=\pi\hbar c/e$ is the magnetic flux quantum.
In these units the Eilenberger equations for quasiclassical
propagators take the form
\begin{align}\label{Eq:EilenbergerF}
&v_{Fj}{\bf n_p}{\bm D} f_j +
 2\omega_n f_j - 2 \Delta_j g_j=0, \\ \nonumber
 &v_{Fj}{\bf n_p}{\bm D}^* f^+_j -
 2\omega_n f^+_j + 2\Delta^*_j g_j=0.
 \end{align}
 Here $v_{Fj}=V_{Fj}/V_{F1}$, $\omega_n=(2n+1)\pi T$ are Matsubara frequencies,
   the vector ${\bf n_p}=(\cos\theta_p,\sin\theta_p)$
  parameterizes the position on 2D cylindrical
 Fermi surfaces. The quasiclassical Green's functions in each band obey
 normalization condition $g_j^2+f_jf_j^+=1$.

 The self-consistency equation for the gaps is
\begin{equation}\label{Eq:SelfConsistentGap}
 \Delta_i=T \sum_{n=0}^{N_d} \int_0^{2\pi}
 \lambda_{ij} f_j d\theta_p.
\end{equation}
 The coupling matrix $\lambda_{ij}$ satisfies the symmetry relations
 $n_1\lambda_{12}=n_2\lambda_{21}$ where $n_i$ are the
 partial densities of states normalized so that $n_1+n_2=1$.
 The vector potential satisfies the Maxwell equation
 $\nabla\times\nabla\times {\bf A} = {\bf j}$ where the current is
 \begin{equation}\label{Eq:SelfConsistentCurrent}
  {\bf j}= -T\sum_{j=1,2} \sigma_j\sum_{n=0}^{N_d}
 Im\int_0^{2\pi}  {\bf n_p} g_j d\theta_p.
\end{equation}
 The parameters $\sigma_j$ are given by
 $ \sigma_j=4\pi\rho n_j v_{Fj}$ and
 $$
 \rho=(2e/c)^2 (r_0 V_{F1})^2\nu_0.
 $$

 The derivation of the TCGL functional (\ref{Gibbs}) from the
microscopic equations\cite{Dao} formally follows the standard
scheme (we present it in the Appendix \ref{Appendix1}). First we
find the solutions of Eqs.(\ref{Eq:EilenbergerF}) in the form of
the expansion by powers of the gap functions amplitudes
$\Delta_{1,2}$ and their gradients. Then these solutions are
substituted to the self-consistency equation for the gap functions
which yields the TCGL theory. The derivation of coefficients in
the expansion (\ref{Gibbs}) is presented in the Appendix
\ref{Appendix1}. Here we denote the values of coefficients
obtained from microscopic theory as
$\bar{a}_\nu,\;\bar{b}_\nu,\;\bar{K}_\nu$ and $\bar{\gamma}$ which
are given by the expressions
\begin{align}\label{Eq:GLexpansionText}
 & \bar{a}_i=  \rho n_i(\bar{\lambda}_{ii}+\ln T -G_c) \\ \nonumber
 & \bar{\gamma}= \rho n_1n_2 \lambda_{J}/{\rm Det}\hat \Lambda \\ \nonumber
 & \bar{b}_i=  \rho n_{i} X/T^2 \\ \nonumber
 & \bar{K}_i=  \rho v_{Fi}^2\bar{b}_i/4
 \end{align}
 where $\lambda_J=\lambda_{21}/n_1=\lambda_{12}/n_2$.
  Here $X=7\zeta(3)/8\pi^2$, ${\bar{\lambda}}_{ij}={\lambda}_{ij}^{-1}$
 and $ G_c=[{\rm Tr} \hat\lambda-\sqrt{{\rm Tr} \hat\lambda^2-4{\rm
 Det}\hat\lambda}]/(2 {\rm Det} \hat\lambda)$.

 Note that in general the derivation of the TCGL model is
{\it not} implemented as an expansion in powers of a single small
parameter $\tau=(1-T/T_c)$ but the outlined above procedure is
based on the assumption of  smallness of  several parameters (gap
functions and their gradients). Indeed   the formal justification
of these assumptions is not straightforward and to the present
moment it has been absent \cite{Dao}. In the present work we show
under what conditions these assumptions are rigorously justified.

\section{Asymptotic behaviour of the fields and coherence lengths.}
\label{Sec:Asymptotic}

First we investigate the asymptotical behaviour of the
 superconducting gaps formulated in terms of the linear modes of the density fields both
in GL \cite{GL2mass} and microscopic \cite{Silaev} theories.
 To find the linear modes we rewrite the equations in terms of the deviations of the gap fields from
their ground state values: $\Delta_i=\Delta_{i0}+\bar{\Delta}_{i}$
where $i=1,2$. To illuminate the qualitatively important physics
we consider a one-dimensional case in the absence of magnetic
field. Let us rewrite the TCGL equations by keeping on the left
hand side  the terms linear in deviations $\bar{\Delta}_{i}$ while
collecting the higher-order nonlinear terms in the r.h.s.:
\begin{eqnarray}\label{GLequations}
  \left[ K_1 d^2/dx^2 - a_1 -3 b_1 \Delta_{10}^2\right]
  \bar{\Delta}_1 +\gamma \bar{\Delta}_2=N_1\\ \nonumber
   \left[ K_2 d^2/dx^2 - a_2 -3 b_2 \Delta_{20}^2\right]
  \bar{\Delta}_2 +\gamma \bar{\Delta}_1=N_2.
\end{eqnarray}
 The r.h.s. gives nonlinear source terms $N_i=b_i(3 \Delta_{i0}
\bar{\Delta}_i^2+\bar{\Delta}_i^3)$.  The solution of
Eq.(\ref{GLequations}) can be found in Fourier representation to
have the form
\begin{equation}\label{Eq:SolFourier}
\bar{\Delta}_i(k)= \hat R^{-1}_{ij} N_j(k)
\end{equation}
 where
 \begin{eqnarray}
  \nonumber & R_{12}=R_{21}=\gamma \\
  \nonumber & R_{ii}=-\left[ K_i k^2 + a_i +3 b_i
  \Delta_{i0}^2\right].
 \end{eqnarray}
 In this case the response function $\hat R^{-1}$ has two poles in the upper
 complex half-plane $k=i\mu_H$ and $k=i\mu_L$
 which determine the {two inverse length scales}
or, equivalently, the two masses of composite gap functions
fields \cite{GL2mass}, which we denote as ``heavy" $\mu_H$ and ``light" $\mu_L$
(i.e. $\mu_H>\mu_L$).

Let us set  $K_1=K_2$ which can be accomplished by rescaling the
fields $\Delta_{1,2}$. Then the matrix $\hat
 R^{-1} (k)$ can be diagonailized with the $k$-independent rotation
  introducing the new linear modes of the fields
 $ {\Phi}_\beta=U_{\beta i} \bar{\Delta}_i$
 and the sources  $N_\beta=U_{\beta i}N_i$ where $\beta=L,H$ and $i=1,2$.
 The rotation matrix $\hat U$ is characterized by
 the mixing angle\cite{GL2mass,Silaev} as
follows:
\begin{equation}\label{Eq:MixingAngle}
\hat U= \begin{pmatrix}
   \cos\theta_L & \sin\theta_L \\
   -\sin\theta_H & \cos\theta_H \
 \end{pmatrix}
  \end{equation}

  Using the diagonal form of the response function
 $\hat R^{-1}(k)$ in the real-space domain we obtain
\begin{equation}\label{Eq:Delta}
{\Phi}_{\beta}(x)= -  \frac{1}{2\mu_\beta} \int_{0}^\infty dx_1
e^{-\mu_{\beta}|x_1-x|} N_{\beta}(x_1) + C_{\beta}
e^{-\mu_{\beta}x}
\end{equation}
where $C_{\beta}=\int_0^\infty
\left[N_{\beta}(x)+2N_{\beta}(0)\right]e^{-\mu_{\beta}x}
dx/2\mu_\beta$ is chosen so that to satisfy the boundary condition
${\Phi}_{\beta}(0)=N_\beta(0)/\mu_\beta^2$ which corresponds to
the condition $\Delta_{1,2}(0)=0$ at $x=0$.

 \subsection{ The limit $\tau \to 0$.}
\begin{figure}[htb!]
\centerline{\includegraphics[width=1.0\linewidth]{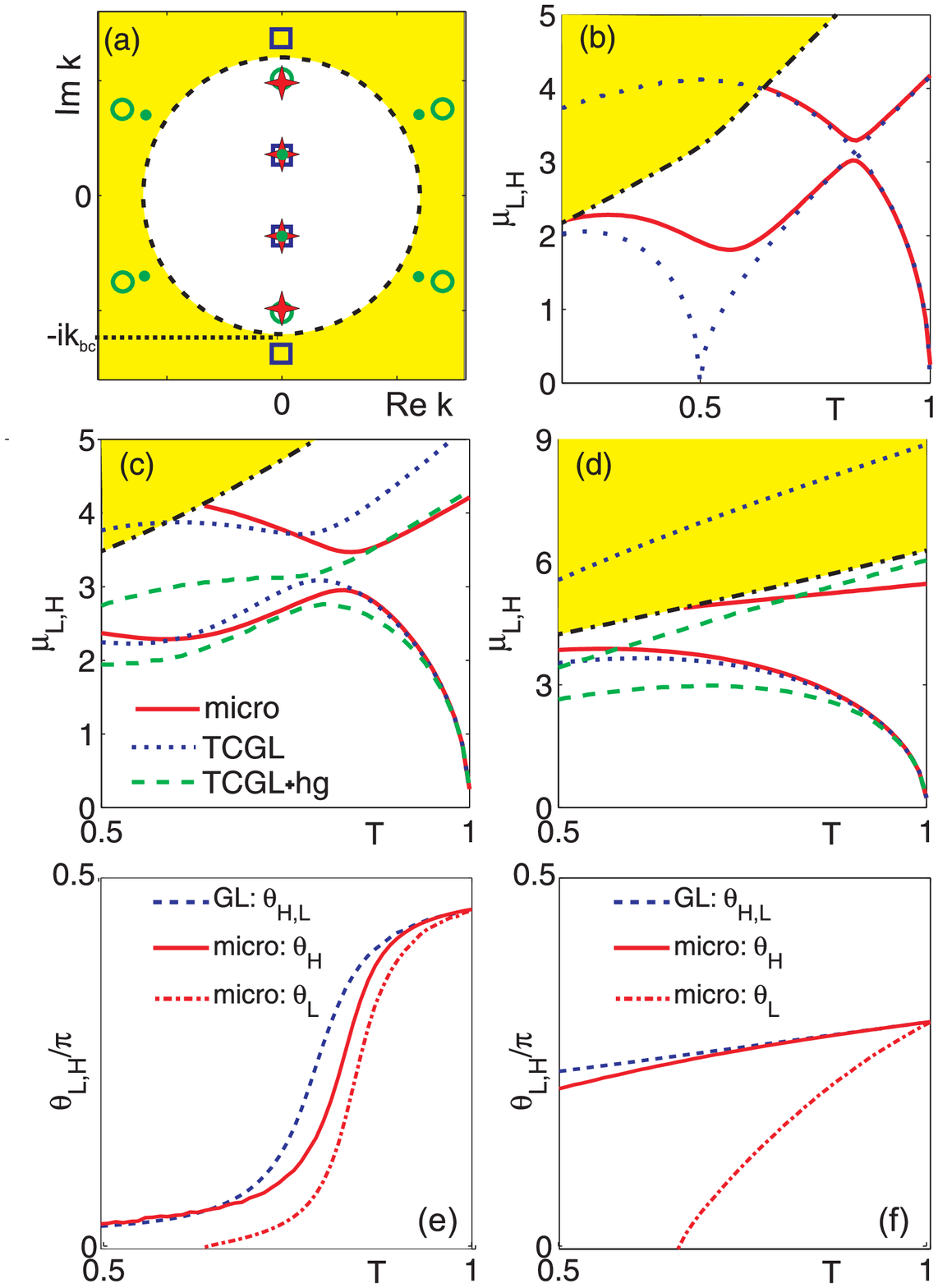}}
\caption{\label{Fig:modes}  (a) Comparison of the response
function singularities in the complex $k$ plane given by the exact
microscopic and microscopically derived TCGL theories. Red crosses
are the physical poles of microscopic theory.  Blue squares
correspond to the conventional TCGL theory while green circles
show the parasitic poles appearing in TCGL expansion up to sixth
order in gradients. The white circle is the area where $\hat R
(k)$ is analytical and $\hat R^{-1} (k)$ is meromorphic. In all
plots (a-d) the yellow shade indicates the area where the response
function is not meromorphic.
 (b) Comparison of the masses of normal modes of
 a $U(1)\times U(1)$ (dotted blue lines) and a
weakly coupled two-band $U(1)$ model  (red lines).  For the weak
component  in the $U(1)\times U(1)$ we also plot the correlation
length for superconducting fluctuations in the normal state for
$T>T_{c2}$. As clearly seen in that plot, adding a Josephson
coupling removes divergence of coherence length of the weak band.
This is because the Josephson coupling represents explicit
symmetry breakdown from $U(1)\times U(1)$ to $U(1)$ and thus
eliminates the phase transition at lower $T_c$. However when this
coupling is weak one of the coherence lengths has a peak around
that temperature. (c) and (d) Comparison of field masses given by
microscopic (solid lines), TCGL (dotted) and TCGL with sixth-order
gradients (dashed lines) theories. The microscopic parameters are
$\lambda_{11}=0.5$, $\lambda_{22}=0.426$  and
$\lambda_{12}=\lambda_{21}=0.005;\;0.01;\;0.1$ for (b,c,d)
correspondingly. (e-f) Comparison of the mixing angle behaviour
given by the exact microscopic (red lines) and microscopically
derived TCGL theories (blue line). Parameters are the same as on
the panels (c-d) correspondingly. Panels (d) and (f) show a
pattern how the TCGL theory  starts to deviate from the
microscopic theory at lower temperature when interband coupling is
increased.}
\end{figure}

 The expression (\ref{Eq:Delta}) shows that two fields
${\Phi}_{L,H}$ vary  at distinct coherence lengths:
$\xi_H=1/\mu_H$ and $\xi_L=1/\mu_L$. They constitute fundamental
length scales of the TCGL theory (\ref{Gibbs}). They  characterize
the asymptotical relaxation of the linear combinations of the
fields $\Delta_{1,2}$, the linear combinations
 are represented by the composite fields ${\Phi}_{L,H}$. Our
calculation shows that these length scales
 behave qualitatively different in the limit $\tau \to 0$.
  Infinitesimally close to $T_c$ the largest length diverges as
$\xi_L\sim \tau^{-1/2}$ while the smaller $\xi_H$ remains finite.
 Similar behavior also follows from full microscopic calculation shown on
Fig.(\ref{Fig:modes})b,c,d where the temperature dependence of
masses $\mu_{L,H}$ is plotted.
  The presence of the non-diverging length scale $\xi_H$  makes the
 qualitative difference with the single-band GL theory but indeed does not contradict the standard
  textbook picture that {\it in the limit} $\tau \to 0$
the mean field theory  of a $U(1)$ system should be well
approximated by single-component GL model. As we show below the
amplitude of the ``heavy" mode vanishes in the $\tau \to 0$ limit
faster than that of ``light" mode. Neglecting the ``heavy" mode
contribution one indeed obtains a single-component GL theory.

We can use the Eq.(\ref{Eq:Delta}) to evaluate
 the asymptotical amplitudes of ${\Phi}_{H,L}(r)$ in terms of
 the powers of the expansion parameter $\tau$, in the limit $\tau \to 0$.
 The goal is to evaluate how the
  contributions from different length scales affect overall profile of the fields as they recover
 their ground state value away from $x=0$. First we note that the source terms $N_{L,H}(x)$
   are confined at the region determined by the
   coherence length $x<\xi_L$. Inside this region the amplitude of the
 deviations of the gaps from the ground state values are large
 so that
 $$
 \bar{\Delta}_i(x) \sim \Delta_{0i}\sim \tau^{1/2}.
 $$
 Thus the amplitude of sources is of the order
$N_{L,H}\sim \tau^{3/2}$. Let us consider the first term in the
expression (\ref{Eq:Delta}) for $\beta=L$ at the asymptotical
region $x>\xi_L$. In this case the integration is confined within
$x_1<\xi_L$ and yields the following estimate ${\Phi}_L\approx A_L
e^{-\mu_L x}$  where
$$
A_L\sim\xi_L\tau^{3/2}\sim\tau^{1/2}.
$$ The
second term in the Eq.(\ref{Eq:Delta}) gives the contribution of
the same order to the amplitude of the ``light" mode.

 The amplitude of the ``heavy" mode is determined entirely by the
first term in Eq.(\ref{Eq:Delta}) for $\beta=H$. We consider the
asymptotical region $ x\ll \xi_L$ therefore in this estimate we
put $N_{H}(x_1)\approx N_{H}(0)$ so that
 $$
 \Phi_{H}(x)=N_{H}(0)\left(2e^{-\mu_H x}-1\right)/\mu_H^2.
 $$
 The function $\Phi_{H}(x)$ has a characteristic scale $\xi_H=1/\mu_H$
 and its overall amplitude is determined by the factor
 $$
 A_H\sim N_H(0)/\mu_H^2 \sim \tau^{3/2}.
 $$
  Thus in the limit $\tau \to 0$ the ``heavy"  mode drops out
because of the vanishing amplitude  $A_H \sim\tau^{3/2}$ as
compared to the ``light" mode $A_L \sim\tau^{1/2}$. Note that
 it is thus principally incorrect to attribute different exponents directly to the functions $\Delta_i$
and to assume that they become equal in the limit $\tau \to 0$ as
claimed in  \cite{kogan} and followed in some other literature
\cite{Shanenko}.


 On the qualitative level we give a less technical but on the
other hand more intuitively transparent description of the
limiting behavior of the fields near $T_c$. We consider the
situation where the coefficients of quadratic terms in
Eq.(\ref{Gibbs}) can be written in the form $a_j(T)=\alpha_j
(T-T_{j})$ with $\alpha_j>0$.
 Thus in the small $\tau$ limit, first the weakest superconducting component becomes  passive: it has
nonzero superfluid density only because of the bilinear Josephson
coupling
 $\gamma (\Delta_1\Delta_2^*+c.c.)$.
To elucidate what happens in the  $\tau \to 0$ limit one can
redefine fields using the following transformation ${\mathbf
\Delta} ={\bf X_D} \Psi_D ({\bf r})+{\bf X_S} \Psi_S({\bf r})$
where ${\bf X_D}=\left(\alpha_2 (T_D-T_2), \gamma\right)^T$,
$T_D\equiv T_c$ is the critical temperature,
 ${\bf X_S}=\left(\gamma, \alpha_1 (T_{S}-T_1)\right)^T$ and $T_S=T_1+T_2-T_D<T_D$.
 This transformation, mixes the gaps and produces a representation where the bilinear Josephson coupling between the new fields
is eliminated at the cost of introducing
 mixed gradient and fourth-order couplings:
 \begin{eqnarray}
 &\nonumber {F} = \sum_{i,j=D,S} K_{ij} \bm D\Psi_{i}\left(\bm D\Psi_{j}\right)^* +
 \beta_i(T-T_{i})|\Psi_i|^2 + \\
 &\sum_{i,j,k,l=D,S} \beta_{ijkl} Re(\Psi_i\Psi_j\Psi^*_k\Psi^*_l).\label{Eq:EnergyPot}
\end{eqnarray}
 In the limit $\tau \to 0$ the dominant component
$\Psi_D$ becomes  single-component GL order parameter while the
component $\Psi_S$ is passive. However instead of being induced by
the bilinear Josephson coupling, $\Psi_S$ is induced by the terms
like $\Psi_S\Psi_D^*|\Psi_D|^2$. Thus for $\tau \to 0$ one has
$\Psi_D\sim \tau^{1/2}$.
  From the fact that $\beta_S(T-T_{S})$ and $\beta_{SDDD}$ are finite
at $\tau = 0$ it follows that in the same limit we have $
\Psi_D\gg\Psi_S\sim \tau^{3/2}$. It means that {\it in the limit
$\tau \to 0$} retaining only the terms containing $\Psi_D$ is
justified and it allows to approximate the model by a conventional
single order parameter theory.

For infinitesimally small $\tau$  there remains only single GL
equation for the order parameter $\Psi_D$
\begin{equation}\label{Eq:SingleComponent}
-K_{DD} {\bf D}^2 \Psi_D+\beta_D (T-T_c)\Psi_D+\beta_{DDDD} \Psi_D
|\Psi_D|^2=0
\end{equation}
where
 $\beta_{DDDD}=(a_{2D}^2 b_1+a_{1D}^2 b_2)/\alpha_1$ and
 \begin{align}\label{Eq:coeffitients}
  &\beta_D=\frac{T_D-T_S}{T_D-T_{2}}>0 \\ \nonumber
  &K_{DD}=\frac{a_{1D}K_2+a_{2D}K_1}{\alpha_1\alpha_2 (T_D-T_2)},
  \end{align}
  where $a_{1,2D}=a_{1,2}(T_D)$.

Thus it is the disappearance of the amplitude of the subdominant
mode which allows one to take a single-component GL limit in this
mean-field theory.  The ``heavy" mode with finite mass even
infinitesimally close to $T_c$ is generated here by the presence
of the term $|\bm D\Psi_S|^2\sim \xi_H^{-2}|\Psi_S|^2$ where
$\xi_H$ is a finite length not diverging at $T_c$.  Therefore near
$T_c$ one has $|\bm D\Psi_S|^2\sim \xi_H^{-2}|\Psi_S|^2\sim
\tau^3$ which is of the same order as the other terms in the free
energy functional. Note that the fields $\Psi_{D,S}$ introduced
here are not directly related to the normal modes of the system
since the mixed gradient and quartic terms will lead to mode
mixing \cite{GL2mass} at finite $\tau$. Also note that despite
there is a growing disparity of the coherence lengths at small but
finite $\tau$ when one approaches critical temperature, it does
not imply that one necessary falls into type-1.5 regime because
$\xi_1 < \lambda <\xi_2$ is only necessary but not sufficient
condition for the appearance of this regime. That is, a system in
some of these cases is type-I despite having $\xi_1 < \lambda
<\xi_2$.


\subsection{GL theory at finite $\tau$.}

 Unfortunately the limiting
$\tau \to 0$ analysis does not have much physical significance in
a generic two-band system. First, the mean-field theory becomes
invalid in the same limit $\tau \to 0$ so the regime where the
system is well described by single-component GL theory can be
cutoff by critical fluctuations. More importantly, as we argue
below, this analysis is in general inapplicable for an assessment
of, e.g. magnetic response of the system. The magnetic response is
a finite-length scale property and requires finite-$\tau$ theory.
Finally,  as shown in microscopic calculations the masses of the
fields in two-band models in certain cases
 change rapidly and in a non-trivial way with decreasing  temperature.
Thus a limiting $\tau \to 0$  analysis in  general cannot give
even an approximate physical picture even at very small $\tau$. In
particular that implies that in a two-band system, a
Ginzburg-Landau parameter (which one, may in principle construct
in the $\tau \to 0$ limit at a mean-field level) is not a useful
characteristic.
 Rather it is required  to make an accurate quantitative study
of two-band theory at finite $\tau$ to determine the conditions
under which the model can be described by singe- or two- component
GL theory or does not allow a description by any such GL
functionals at all. In order to do it we utilize the exact form of
the response function $\hat R^{-1}$ (i.e. valid at any $T$) found
from the linearized microscopic theory according to the procedure
developed in \cite{Silaev}. In contrast to the GL theory, the
microscopic response function $\hat R^{-1} (k)$ has branch cuts
along the imaginary axis starting at point $\pm ik_{bc}$ where
$k_{bc}=2\sqrt{\Delta_{02}^2+(\pi T)^2}$ (we assume that
$\Delta_{02}<\Delta_{01}$).
   Inside the circle $|k|<|k_{bc}|$ shown by the white area
 in Fig.(\ref{Fig:modes})a the response function is meromorphic, i.e. its singular points are
 only poles shown by the red crosses. The non-meromorphic region is
 marked by the yellow shade in all panels of
 Fig.(\ref{Fig:modes}).
  In general inside the meromorphic circle there can be two poles of
$\hat R^{-1} (k)$ at $Im(k)>0$ shown by red crosses in
Fig.(\ref{Fig:modes})a. Analogously to TCGL these poles determine
the masses $\mu_{L,H}$ of the ``heavy" and ``light" modes, and
thus the corresponding coherence lengths. The contribution of the
branch cut contains the continuous spectrum of length scales
shorter than $1/k_{bc}$ which can not be described within GL
theory. Moreover for some parameters (e.g. at strong Josephson
coupling) one of the poles which corresponds to the ``heavy" mode
can disappear by merging with the branch cut. In this case there
is only one fundamental length scale left since the contribution
of the ``heavy" mode can not be separated from the branch cut.

The microscopically calculated temperature dependencies of masses
of the modes in a superconductor with weak interband
coupling are shown in Fig.(\ref{Fig:modes})b by red solid lines.
For  a reference we also plot masses in $U(1)\times U(1)$ theory
which has two independently diverging coherence lengths at
$T=T_{c1}$ and $T=T_{c2}$ (chosen to be $T_{c2}=0.5T_{c1}$). For
coupled bands the hybridization of modes removes the divergence at
$T=T_{c2}$ and introduces the avoided crossing point of the
``heavy" and ``light" modes.

Let us now assess the applicability of minimal TCGL model
Eq.(\ref{Gibbs}) without using expansion in powers of $\tau$.
Compared to the previous works \cite{Dao}, we use more complicated
temperature dependence of the coefficients derived in the Appendix
\ref{Appendix1}. Let us compare  the behavior of the masses of the
modes in the microscopically derived TCGL and a full microscopic
theory. It is shown for the cases of weak and strong interband
coupling in Fig.(\ref{Fig:modes})c,d. We have found that  TCGL
theory describes the lowest characteristic mass $\mu_{L}(T)$ with
a very good accuracy near $T_c$ [compare the blue and red curves
in Fig. (\ref{Fig:modes})c,d]. Remarkably, when interband coupling
is relatively weak [Fig.(\ref{Fig:modes})c] the ``light" mode is
quite well described by TCGL also at low temperatures down to
$T=0.5 T_c$ around which the weak band crosses over from active to
passive (proximity-induced) superconductivity. Indeed the $\tau$
parameter is large in that case and cannot be used at all to
justify a GL expansion. Nonetheless if the interband coupling is
small one does have a small parameter to implement a GL expansion
for one of the components. Namely one can still expand, e.g. in
the powers of the weak gap $|\Delta_2|/\pi T\ll 1$.
 On the other hand for the
``heavy" mode we obtain some discrepancies even relatively close to $T_c$,
although TCGL theory gives qualitatively correct picture for this
mode when the interband coupling is not too strong. More
substantial discrepancies between TCGL and microscopic theories
appear only at lower temperatures or at stronger interband
coupling [Fig.(\ref{Fig:modes})d] where the microscopic response
function has only one pole, while TCGL theory generically has two
poles.

\begin{figure}[h!]
\includegraphics[width=1.0\linewidth]{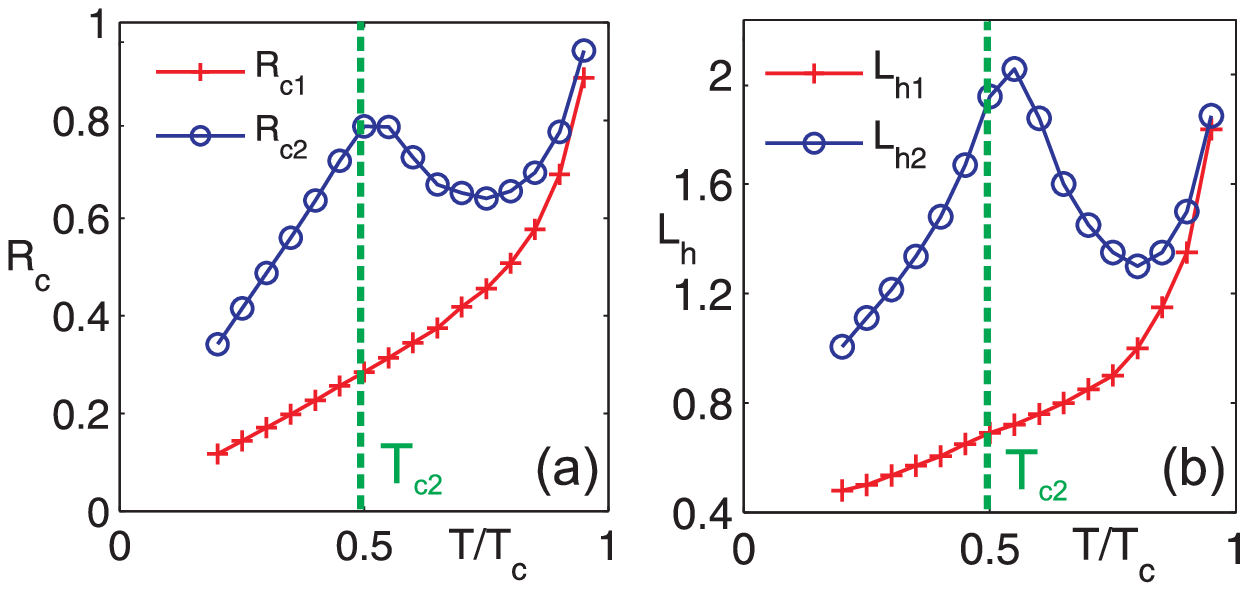}
\caption{\label{Fig:CoreSize} (a) Sizes of the vortex cores
$R_{c1,2}$ and (b) healing lengths $L_{h1,2}$ in weak (blue curve,
open circles) and strong bands (red curve, crosses) as functions
of temperature. The parameters are $\lambda_{11}=0.5$,
$\lambda_{22}=0.426$, $\lambda_{12}=\lambda_{21}=0.0025$ and
$v_{F2}/v_{F1}=1$. In the low temperature domain, the vortex core
size in the weak component grows and reaches a local maximum near
the temperature $T_{c2}$ (the temperature near which the weaker
band crosses over from being active to having superconductivity
induced by an interband proximity effect) \cite{Silaev}. In the
absence of interband coupling there is a genuine second
superconducting phase transition at $T_{c2}  = 0.5 T_{c1}$ where
the size of the second core diverges. When interband coupling is
present  it gives an upper bound to the core size in this
temperature domain, nonetheless this regime is especially
favorable for appearance of type-1.5 superconductivity
\cite{Silaev}. }
\end{figure}

Comparison of the masses of normal modes of
 a $U(1)\times U(1)$ (dotted blue lines) and a
weakly coupled two-band $U(1)$ model  (red lines) shown in
Fig.(\ref{Fig:modes})d demonstrates that adding a Josephson
coupling removes divergence of coherence length of the weak band.
This is because the Josephson coupling represents explicit
symmetry breakdown from $U(1)\times U(1)$ to $U(1)$ and thus
eliminates one of the superconducting  phase transitions at lower $T_c$. However when this
coupling is weak one of the coherence lengths has a substantial peak around
that temperature. The peaked behaviour of coherence length near
the critical temperature of the weak superconducting band is has
clear physical manifestation in the temperature dependence of the
vortex core size. Let us note that to assess the overall size of
core requires analysis of full nonlinear theory. In Fig.(\ref{Fig:CoreSize}) we plot the sizes of
the vortex cores in weak and strong bands calculated in the full
nonlinear model according to the two alternative definitions. The
first one is the slope of the gap function distribution at $r=0$
which characterizes the width of the vortex core near the center
$R_{cj}=(d\ln\Delta_{j}/dr)^{-1}(r=0)$
[Fig.(\ref{Fig:CoreSize})a]. The second one is the healing length
$L_{hj}$ defined as $\Delta_{j}(L_{hj})=0.95\Delta_{0j}$
[Fig.(\ref{Fig:CoreSize})b] (i.e. this length is not directly related to exponents
but quantifies at what length scales the gap functions almost recover
their ground state values). Both definitions demonstrate the
stretching of the vortex core in the weak component related to the
peak of the coherence length shown in the Fig.(\ref{Fig:modes})d.
Note that the weak band healing length $L_{h2}(T)$ in
Fig.(\ref{Fig:CoreSize})b has maximum at the temperature slightly
larger than $T_{c2}$ which is consistent with the fact that the
maximum of coherence length $\xi_L$ (equivalently the minimum of
the field mass $\mu_L$) in Fig.(\ref{Fig:modes})d is shifted to
the temperature above $T_{c2}$  ($T_{c2}$ is defined as the lower critical temperature in the
limit of no Josephson coupling).

\subsection{Effects of higher order gradient terms.}

The origin of the small disagreement between the TCGL and
microscopic masses of the ``heavy" mode is the absence of
higher-order gradient terms in the expansion (\ref{Gibbs}). The
inclusion of higher order gradients means adding more terms to the
Taylor expansion of the function $\hat R (k)$ which is known to
converge inside the circle $|k|<|k_{bc}|$. Using this procedure
one can get  a better agreement with microscopic theory (compare
green dashed and red solid lines in Fig.(\ref{Fig:modes})c,d).
However,  such an extension is hardly useful because as a
byproduct it generates unphysical artifacts such as many parasitic
poles of the response function $\hat R^{-1} (k)$ outside the
circle $|k|<|k_{bc}|$. The position of parasitic poles appearing
in the sixth-order gradient expansion is shown by green circles in
the Fig.(\ref{Fig:modes}a). These parasitic poles lie outside the
imaginary axis, thus yielding unphysical oscillating contributions
to asymptotical behavior of the corresponding linear modes.

\subsection{Characteristic length scale of the phase difference variations.}
 In the $U(1)\times U(1)$ system one has two massive modes
associated with the  modules of the complex fields and a Goldstone
boson associated with the phase difference. If one adds a
Josephson coupling there appears also the third mass parameter.
When Josephson term is present the phase difference acquires a
preferred value. Its deviations from the preferred value are
characterized by a mass parameter. In the constant density
approximation the terms of the TCGL functional which describe the
phase difference mode are:
 \begin{equation}
\frac{1}{2}\frac{K_1K_2\Delta_{10}^2
\Delta_{20}^2}{K_1\Delta_{10}^2+K_2\Delta_{20}^2} (\nabla
(\theta_1 -\theta_2))^2 -\gamma \Delta_{10}\Delta_{20} \cos
(\theta_1 -\theta_2)
 \end{equation}
Expanding the second term gives  the mass parameter for the phase
difference mode
 \begin{eqnarray}
 m_{(\theta_1-\theta_2)}=\sqrt{{\gamma}\frac{K_1\Delta_{10}^2+K_2\Delta_{20}^2}
 {K_1K_2\Delta_{10}\Delta_{20}} }.
 \end{eqnarray}
It is useful to consider the behavior of this mass in the limit $T
\to T_c$. In that case we have $\Delta_{10,20} \propto
\sqrt{1-T/T_c}$. Thus
 \begin{eqnarray}
\lim_{T \to T_c}(m_{(\theta_1 -\theta_2)}) \to const
 \end{eqnarray}

 To summarize this part we have shown that the
  TCGL model of the form given by Eq. (\ref{Gibbs}) with microscopically
derived temperature dependencies of coefficients is overall highly
accurate at elevated temperatures.
 The small discrepancies with microscopic theory affect only short-length
scale physics which implies that TCGL model gives the precise
answer for long-range intervortex forces. Also we find that in
some cases the TCGL model provides an accurate description of the
large length scales physics at temperatures much lower than $T_c$.
In Appendix \ref{App:GLexpansion}  we discuss the origin of the
disagreements between these results and some of the recent
literature \cite{Shanenko}.

 \section{Vortex structure: TCGL vs microscopic theory.}

For inhomogeneous situations, such as vortex solutions,
the overall profiles of the fields is affected not only
by fundamental length scales (i.e. coherence lengths) but also by nonlinear effects.

 Let us now study non-linear effects case of vortex solutions. Obviously, because of the growing
importance of nonlinear effects at lower temperatures the
Eq.(\ref{Gibbs}) cannot describe quantitatively well the total
structure of vortices when $T\ll T_c$. In
Fig.(\ref{Fig:Comparison}) we compare the vortex solutions in
 the self-consistent microscopic theory (red dotted
curves) and in the corresponding TCGL theory with coefficients
obtained by expansion (blue dashed-dotted curves). One can see
that at elevated temperatures the agreement is very good but for
lower temperatures there is a growing discrepancy. One of the
reasons behind the discrepancy is the trivial shift of the ground
state values of the fields by nonlinearities. Note that  at the
level of GL theory the inclusion of more nonlinear terms merely
renormalizes masses and length scales but does not alter the form
of linear theory\cite{GL2mass}. {Thus in the current example of
the full nonlinear model it is also reasonable to check if one
could get a better agreement with microscopic theory by treating
the coefficients in the minimal TCGL model Eq.(\ref{Gibbs}) {
phenomenologically.  For all practical purposes this provides
alternative route to the more restrictive approach of finding a
refined microscopic expansion}.  A good agreement with the
microscopic theory in this procedure will imply that the system
does posses a description in terms of a classical two-component
field theory.


We compared the vortex solutions in the TCGL theory with fitted
coefficients and the exact microscopic model for the particular
example of the system with coupling constants $\lambda_{11}=0.5$,
$\lambda_{22}=0.46$, $\lambda_{12}=\lambda_{21}=0.005$ and
$v_{F2}/v_{F1}=5$. The values of the coefficients which provide
the best fit are listed in the table \ref{table1}.
\begin{table}
\begin{center}
\begin{tabular}{|p{1cm}||p{1cm}|p{1cm}|p{1cm}|p{1cm}|p{1cm}|p{1cm}|p{1cm}|}
 \hline
 $T$ & $0.98$ &$0.8$ &$0.7$ & $0.5$& $0.4$& $0.2$
  \\ \hline\hline
 $b_1$  &$0.95\bar{b}_1$ &$0.85\bar{b}_1$ &$0.76\bar{b}_1$ &$0.54\bar{b}_1$ &$0.44\bar{b}_1$ &$0.18\bar{b}_1$
  \\ \hline
 $K_1$ &$\bar{K}_1$ &$0.8\bar{K}_1$ &$0.65\bar{K}_1$ &$0.5\bar{K}_1$ &$0.4\bar{K}_1$ &$0.15\bar{K}_1$
 \\ \hline
 $b_2$  &$\bar{b}_2$ &$\bar{b}_2$ &$\bar{b}_2$ &$0.76\bar{b}_2$ &$0.64\bar{b}_2$ &$0.29\bar{b}_2$
 \\ \hline
  $K_2$  &$\bar{K}_2$ &$0.55\bar{K}_2$ &$0.35\bar{K}_2$  &$0.3\bar{K}_2$  &$0.15\bar{K}_2$ &$0.08\bar{K}_2$
  \\ \hline
 \end{tabular}
 \caption[]{Fitting of TCGL coefficients to match the solutions of exact microscopic equations.
 Here we denote the values of
coefficients obtained from microscopic theory as
$\bar{a}_\nu,\;\bar{b}_\nu,\;\bar{K}_\nu$ and $\bar{\gamma}$.}
\label{table1}
 \end{center}
 \end{table}

By the green dashed lines we show the fits obtained by setting the
values of the TCGL coefficients as listed in the table
\ref{table1}. By doing it we find that
 the solutions of microscopic equations
in a large region of parameters can be fitted with excellent
accuracy by the effective TCGL theory, even at quite low
temperatures. The main discrepancies at very low temperatures
arise due to non-local effects which lead to the disappearance of
the ``heavy" asymptotic mode as well as due to Kramer-Pesch-like vortex
core shrinking\cite{Silaev,agterberg} in both components  which
cannot be captured in the TCGL field theory. Note however that
even in the case where TCGL description starts breaking down, the
discrepancy is mostly pronounced near the origin of the core,
while the soft modes and long-range intervortex interaction can be
well described by a phenomenological TCGL theory.

\begin{figure}[h!]
\includegraphics[width=1.0\linewidth]{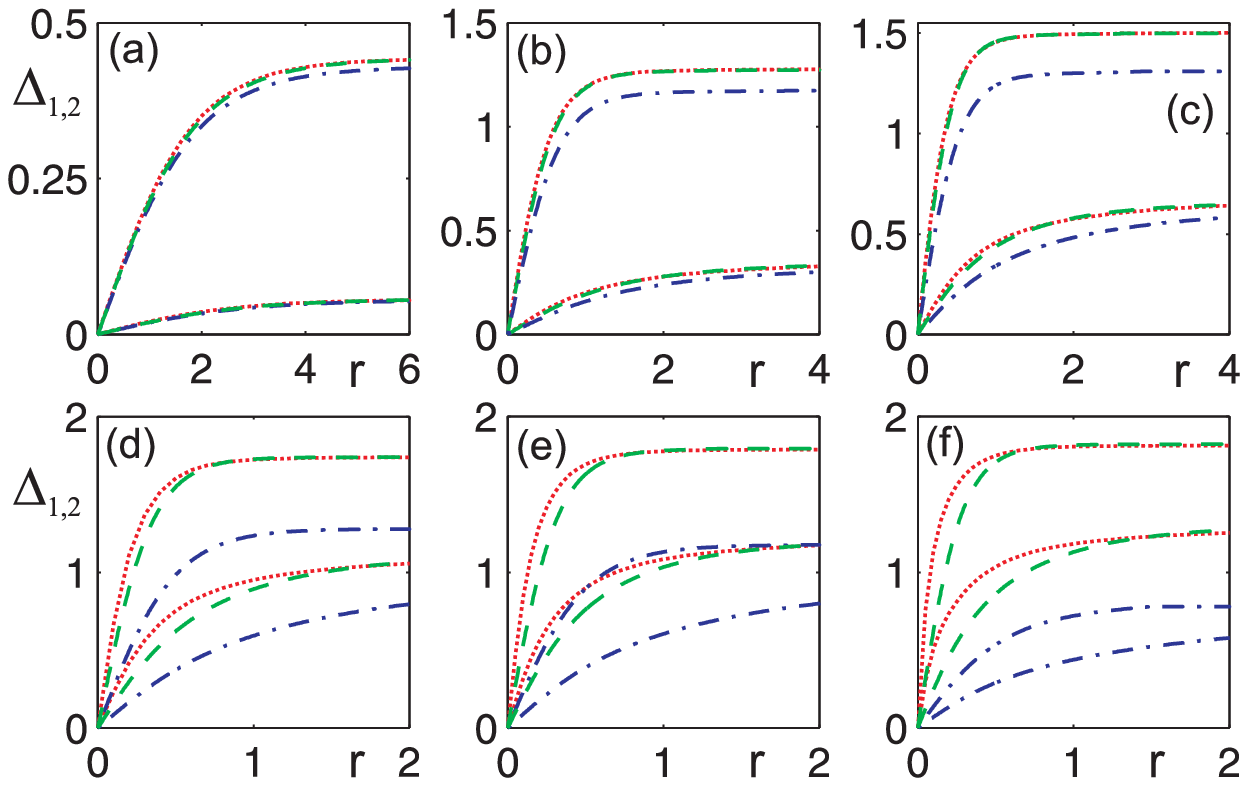}
\caption{\label{Fig:Comparison}
Behavior of the gap functions $\Delta_{1,2}(r)$ in a vortex solution.
  Comparison of the results of
exact microscopic calculation (red dotted lines), TCGL with
microscopically calculated coefficients  (blue dash-dotted lines) and TCGL with
phenomenologically fitted
coefficients (green dashed lines)  at (a) $T=0.98$, (b) $T=0.8$,
(c) $T=0.7$, (d) $T=0.5$, (e) $T=0.4$, (f) $T=0.2$. Coupling
constants are $\lambda_{11}=0.5$, $\lambda_{22}=0.46$,
$\lambda_{12}=\lambda_{21}=0.005$ and $v_{F2}/v_{F1}=5$.}
\end{figure}

\section{Conclusions}

 TCGL model is widely used for  describing  various
aspects of multiband superconductivity. However the TCGL expansion
has never been rigorously   justified for two-band systems, and the current literature contains
diametrically opposite claims regarding the validity of the expansion
or such  basic aspects as the form of  TCGL
functional and behavior of the coherence lengths near $T_c$ \cite{Dao,kogan,Shanenko}.
We investigated under which conditions a two-band system can be
described by a  TCGL theory. First we obtained  a TCGL model with a microscopically derived temperature dependence
of coefficients (more general than what could be  obtained in a
straightforward $\tau$ expansion) and demonstrated that it
 gives an accurate description of length scales
and vortex solutions at elevated temperatures
by a comparison with an exact microscopic theory. Second we have
shown that, in a much wider range of temperatures,
 the minimal TCGL model with phenomenologically adjusted
coefficients  gives an accurate description of linear and
nonlinear physics such as vortex excitations and thus the magnetic
response of the system.

The existence of two coherence lengths  $\xi_L,\xi_H$ along with
the magnetic field penetration lengths $\lambda$ in the TCGL model
makes it impossible in general to define the Ginzburg-Landau
parameter in two-band systems, unless one takes the limit  $\tau
\to 0$. In contrast to $U(1)\times U(1)$ superconductors, the
two-band systems have only $U(1)$ symmetry, and as we discussed
above it guarantees that one of the modes drops out of the mean
field theory the limit  $\tau \to 0$ allowing one to define in
that limit $\kappa_{GL}=\xi_L (T\to T_c)/\lambda (T\to T_c)$.
However, even slightly away from the limit   $\tau \to 0$, when
interband coupling is weak the ratio $\xi_L/\lambda$ has a very
strong temperature dependence and the second mode develops with
the coherence $\xi_H$. Thus in general $\kappa_{GL}$ cannot be
used as universal characteristic of magnetic response of two-band
systems.

This work is supported by Knut and Alice Wallenberg Foundation through the Royal Swedish Academy of Sciences, Swedish Research Council,
 US NSF CAREER Award No. DMR-0955902,
"Dynasty'' foundation, Presidential RSS Council (Grant No. MK-4211.2011.2) and Russian Foundation for Basic
Research.

 \appendix

\section{Microscopic model and derivation of TCGL}
\label{Appendix1}

\subsection{Ginzburg-Landau expansion.}
\label{App:GLexpansion}

 To derive differential GL equations (\ref{GLequations}) from the microscopic theory, first  we
find the solutions of Eilenberger Eqs.(\ref{Eq:EilenbergerF}) in
the form of the expansion by the gap functions amplitudes
$|\Delta_{1,2}|$ and their gradients $|({\bm D \bf n_p})
\Delta_{1,2}|$. Then these solutions are substituted to the
self-consistency Eq.(\ref{Eq:SelfConsistentGap}). Using this
procedure we find the solutions of Eqs.(\ref{Eq:EilenbergerF}) in
the form:
\begin{equation}\label{Eq:2OrderExpansion}
f=\frac{\Delta}{\omega_n}-\frac{|\Delta|^2\Delta}{2\omega_n^3}-\frac{v_F}{2\omega_n^2}
({\bm D \bf n_p}) \Delta+\frac{v_F^2}{4\omega_n^3} ({\bm D \bf
n_p}) ({\bm D \bf n_p}) \Delta.
\end{equation}
and $f^+({\bf n_p})=f^*(-{\bf n_p})$. Note that a GL expansion is
based on neglecting the higher-order terms in powers of $|\Delta|$
and $|({\bm D \bf n_p}) \Delta|$. Indeed this approximation
naturally fails in in a number of cases. In this work we determine
the regimes when it can be justified, in particular  by direct
comparison with exact microscopic model. Let us determine
microscopic coefficients in the GL expansion.
Substituting to the self-consistency
Eqs.(\ref{Eq:SelfConsistentGap}) and integrating by $\theta_p$ we
obtain
\begin{eqnarray}\label{}
  \Delta_1=(\lambda_{11} \Delta_1+\lambda_{12} \Delta_2)  G +  (\lambda_{11} GL_1+\lambda_{12} GL_2) \\
    \Delta_2=(\lambda_{21} \Delta_1+\lambda_{22} \Delta_2)  G +  (\lambda_{21} GL_1+\lambda_{22} GL_2)
\end{eqnarray}
where
\begin{equation}\label{Eq:GX}
 G=2\sum_{n=0}^{N_d} \frac{\pi T}{\omega_n};\;\;\;
X=\sum_{n=0} \frac{\pi T}{\omega_n^3}
\end{equation}
\begin{equation}\label{}
  GL_j=X\left(\frac{v_{Fj}^2}{4} {\bm D}^2 \Delta_j- |\Delta_j|^2\Delta_j\right) \\
\end{equation}
Expressing $GL_i$ from the equations above we obtain
\begin{eqnarray}
 n_1 GL_1=n_1 \left(\frac{\lambda_{22}}{{\rm Det} \hat\Lambda}-G\right) \Delta_1-
 \frac{\lambda_{J}n_1n_2}{{\rm Det} \hat\Lambda} \Delta_2 \\
 n_2 GL_2=n_2\left(\frac{\lambda_{11}}{{\rm Det} \hat\Lambda}-G\right) \Delta_2-
 \frac{\lambda_{J}n_1n_2}{{\rm Det} \hat\Lambda} \Delta_1
\end{eqnarray}
Comparing these Eqs with Eqs (\ref{GLequations}) we obtain the
expression for the coefficients
\begin{align}\label{Eq:GLexpansion}
 & \bar{a}_i=  \rho n_i(\tilde{\lambda}_{ii}+\ln T -G_c) \\ \nonumber
 & \bar{\gamma}= \rho n_1n_2 \lambda_{J}/{\rm Det}\hat \Lambda \\ \nonumber
 & \bar{b}_i=  \rho n_{i} X/T^2 \\ \nonumber
 & \bar{K}_i=  \rho v_{Fi}^2\bar{b}_i/4
 \end{align}
 where $\lambda_J=\lambda_{21}/n_1=\lambda_{12}/n_2$.
  The temperature is normalized to the $T_c$. Here
$X=7\zeta(3)/8\pi^2$, ${\bar{\lambda}}_{ij}={\lambda}_{ij}^{-1}$
and
 $$
 G_c=G(T_c)=\frac{{\rm Tr} \lambda-\sqrt{{\rm Tr} \lambda^2-4{\rm Det} \lambda}}{2 {\rm Det}
 \lambda}.
 $$
 We have used the expression $G(T)=G(T_c)-\ln
 T$. Near the critical temperature $\ln T \approx -\tau$ and we obtain
$\bar{a}_i= n_i\lambda_{J}(T-T_i)$ where
$T_i=(1+G_c-\tilde{\lambda}_{ii})$.

\subsection{Remark on $\tau$-expansion}

 Here we comment on the origin of the
qualitative disagreement of our result compared to the
work \cite{Shanenko} which aims at calculating higher order corrections in $\tau=(1-T/T_c)$.
Fist let us make a few general remarks: Note that in the
 derivation of TCGL theory we do not implement an expansion
in powers of $\tau=(1-T/T_c)$. Instead we retain  more complicated
temperature dependence of the coefficients. Also we stress that
any approach to GL expansion depends on what parameters are
assumed to be small, the question is always how and for what
parameters such an assumption is justified.
 The origin of the principal difference
in the behavior of the length scales (Ref. \cite{Shanenko} asserts
that there are two divergent length scales when $\tau \to 0$).
It originates in the
 adoption in   \cite{Shanenko}  of a $U(1)\times U(1)$ theory
as the leading order in expansion following the erroneous
derivation in \cite{kogan} (see the discussion of the errors in
that derivation in \cite{comment1}). Another problem with {a
straightforward implementation} of the expansion by $\tau$ is that
{\it in general} it is incontrollable in the next to leading order
in two-band theories, if one explicitly retains two gap fields.
This is because in contrast to the single-component GL theory, in
general it is not possible to classify different terms by powers
of the parameter $\tau$. As shown in the main body of the paper,
system contains a mode with non-diverging coherence length so that
the spatial derivatives in general do not necessary add the power
of $\tau$. Also since the work \cite{Shanenko} uses as a leading
order the incorrect derivation from \cite{kogan} it  requires
adjustments.

\section{Asymptotic in two-dimensional vortex problem}

The consideration of asymptotic modes in Sec.\ref{Sec:Asymptotic}
can be generalized for the two-dimensional axially symmetric
problem, which allows to treat the asymptotical behaviour of the
gap functions far from the vortex core. First of all in this case
one should substitute the $d^2/dx^2$ by $\nabla^2_r=
d^2/dr^2+r^{-1} d/dr$ in Eq. (\ref{GLequations}).

 Choosing the proper value of the mixing angle the
l.h.s. of Eq. (\ref{GLequations}) can be diagonalized and the
system acquires the form
\begin{equation}\label{GLequations1}
  \left( \bm \nabla^2_r - \mu_i^2 \right)
  \bar{\Psi}_i = N_i
\end{equation}
where the nonlinear $N_{H,L}$ are obtained according to the rule
(\ref{Eq:MixingAngle}).

 Our interest is the asymptotical behaviour of the
 fields $\bar{\Delta}_{L,H}$ determined by the equation above.
 The solution of Eq.(\ref{GLequations1}) can be found in Fourier representation
  $\bar{\Delta} (k)= \int_{-\infty}^{\infty} \bar{\Delta} (x) e^{ikx} dx$ to have the form
 (\ref{Eq:SolFourier}). In this particular case the response function is a
 diagonal matrix:
 $$
 \hat R(k)=\begin{pmatrix}
 (k^2+\mu_H^2) & 0\\
 0 & (k^2+\mu_L^2)\
 \end{pmatrix}
 $$
 In the real-space  domain the field components can be
 expressed with the help of Fourier-Bessel transform
 $$
 \bar{\Psi}_i=\int_0^{\infty}
 J_0(kr_1)J_0(kr) \hat R^{-1}_{ij} N_j(r_1) kdk r_1dr_1.
 $$
 The integration by $k$ in this expression can be performed by
 transforming the contour in the complex plane.
 Using the exact form of the response function the fields asymptotic is found to be given
 by the following expression
\begin{eqnarray}\label{Eq:DeltaR}
\bar{\Psi}_i(r)=\pi K_0(\mu_i r)
 \int_0^r r_1 dr_1 I_0(\mu_i r_1) N_i(r_1)+\\ \nonumber
 \pi I_0(\mu_i r)
 \int_r^{\infty} r_1 dr_1 K_0(\mu_i r_1) N_i(r_1)
\end{eqnarray}
 where $K_0$ and $I_0$ are modified Bessel functions having the
 following asymptotics $K_0,I_0(x)\approx e^{\mp x}/\sqrt{x}$.

 The expression (\ref{Eq:DeltaR}) yields a number of length scales
 characterizing the asymptotical relaxation of the gap fields. The largest
 length is the mean field coherence length $\xi_L=1/\mu_{L}\sim
 1/\tau^{1/2}$. However the presence of the another linear mode in the
 theory sets the scale which is proportional to $\xi_H=1/\mu_{H}$.
This scale remains finite even at $T=T_c$ but its amplitude
vanishes.

\end{document}